\documentclass{aa}  

\usepackage{graphicx}
\usepackage{txfonts}
\usepackage{lipsum}
\usepackage{subcaption}         
\usepackage{lscape}             
\usepackage{placeins}           
\usepackage{booktabs}
\usepackage{CJKutf8}
\usepackage[switch]{lineno}
\usepackage{color}
\usepackage[colorlinks, citecolor=blue]{hyperref}

\begin{document}

   \title{Is gamma-ray burst polarization from photosphere emission?}



   \author{Yan-Zhi Meng\inst{1}, Shu-Qing Zhong\inst{1}, Jia-Hong Gu\inst{1}, Xin-Fei Li \inst{1}  \and Xiaozhou Zhao (\begin{CJK*}{UTF8}{gbsn}赵小舟\end{CJK*})\inst{2}
        }

   \institute{School of Science, Guangxi University of Science and Technology, Liuzhou, Guangxi 545006, People's Republic of China\\
             \email{yzmeng2023@126.com}
              \and Yunnan Observatories, Chinese Academy of Sciences, Kunming,  People's Republic of China\\ }

   \date{Received September 30, 20XX}

 
  \abstract
   {Despite more than half a century of research, the dominant radiation mechanism of gamma-ray burst (GRB) prompt emission remains unsolved. Some progress has been made through the analyses of the observational spectra of  \textit{Swift}/BAT, Konus/Wind, and \textit{Fermi}/GBM, as well as the spectra of the photosphere or synchrotron models, but it is still insufficient to pin down the answer. }
   {Combining the spectral and polarization observations, we seek new criteria for model evaluation.} 
   {We thoughtfully investigate the polarization samples of POLAR and AstroSAT, combining the light curve, the spectral, and the polarization parameters.}
   {The power-law shape of the X-ray afterglows, the $T_{90} \propto (L_{\text{iso}})^{-0.5}$ correlation, and the hard low-energy spectral index $\alpha$ are revealed, thus supporting the photosphere origin. Furthermore, we discover the positive correlation of the $\alpha$ and the polarization degree (PD), which can be consistently explained by the photosphere polarization scenario involving the jet asymmetry from a moderate viewing angle of $\theta _{v}$=0.015.}
   {}

  \authorrunning{Yan-Zhi Meng et al. (2025)}
  \maketitle


\section{Introduction}

\label{sec:Intro}

Gamma-ray bursts are the strongest explosions in our universe.
Several thousands of timing and spectral observations have been accumulated
since its first discovery in the late 1960s. Based on these, the
ultra-relativistic jet is believed responsible for producing GRBs.
However, the jet properties (especially the structure and composition) are
still not well known \citep{Me2002,Zhang2020}. With additional observation information, the azimuthal scattering angle distribution of the incoming photons (standing for the polarization), these properties could be better constrained. The azimuthal
scattering angle is defined as the angle between a fixed X-axis (in the X-Y plane perpendicular to the moving direction of the incoming photon) and the projection of the momentum vector of the outgoing photon (after Compton scattering in the first detector segment) onto the X-Y plane. To obtain this azimuthal angle, the outgoing photon should be detected to interact with a second detector segment, through a second Compton scattering or the photo-absorption. The azimuthal scattering angle is then deduced from the position of these two detector segments. Finally, a histogram of the azimuthal scattering angle (namely the azimuthal scattering angle distribution) for all the photons performing more than two interactions is achieved, referred to as a modulation curve. The amplitude of this curve represents the polarization degree and its phase is related to the polarization angle (PA). Up until now, only about two dozen  polarization detections have been reported \citep{Cobur2003,Yone2011,Yone2012,Covi2016,Chat2019,Zhang2019,Kole2020,Chat2022},
due to the difficulty of detection. Meanwhile, many detections suffer from
substantial uncertainties, mainly the systematic uncertainties \citep{Gill2021}.
Recently, the results of the dedicated GRB polarimeter, POLAR, suggest
that the GRB polarization degree is around $10\%$ \citep{Zhang2019}, which is lower than the prediction of many GRB models, especially those involving synchrotron radiation \citep{Granot2003,Lyuti2003,Waxman2003}. Therefore, the photosphere emission model may be more consistent with these results \citep{Lund2014}.

The photospheric emission is the prediction of the original fireball model %
\citep{Good1986,Pacz1986}, owing to that the optical depth $\tau $ at the outflow
base far exceeds unity \citep[e.g.,][]{Pi1999}. As the fireball
expands and the optical depth declines, the internally trapped photons
ultimately escape at the photosphere radius $R_{\text{ph}}$ ($\tau =1$; \citealt{Rees2005,Pe'er2008,Belo2011,Ruffini2013}). Indeed, based on the
analyses of the observed spectral shape, a quasi-thermal component has been
discovered in a great deal of BATSE GRBs \citep{Ryde2009} and
numerous \textit{Fermi} GRBs (especially in GRB 090902B; \citealt{Abdo2009,Ryde2010,Pe'er2012}). Also, some
statistical aspects of the spectral analysis results for large GRB sample \citep[e.g.,][]{Acuner2020,Dere2020,Gowri2024} seem
to support that the typical observed Band function (smoothly joint broken
power law, \citealt{Band1993}) or cutoff power law can be explained by the photosphere emission,
namely the photospheric emission model. First, many of observed bursts have a harder low-energy spectral index than the death line $\alpha $= $-$2/3 of
the basic synchrotron model, especially for the peak-flux spectrum and short
GRBs \citep[e.g.,][]{Kan2006,ZhaBB2011}. Second, the cutoff power law is the best-fit spectral model for more than half of the GRBs. Thus, the exponential decay in the high-energy end is more consistent with the photosphere emission model. The soft low-energy spectrum can be reconciled by considering the overlapping of blackbodies with various temperatures emitted from different positions (the probability photosphere model or the nondissipative photosphere model; \citealt{Pe'er2008}). Third, for a large proportion of GRBs, the spectral width is found fairly narrow \citep{AxBo2015}\footnote{Notice that, as stated in \citet{Burgess2019}, some narrow spectra can also be fitted by the physical synchrotron model. The width measure can suffer from the poor fit of the empirical Band function, and the smoothly-broken power law (SBPL) model with a variable curvature parameter $\Delta$ could be a better model.}. Fourth, in \citet{Meng2022}, by separating the GRB sample into three sub-samples according to the prompt efficiency $\epsilon _{\gamma}$ ($\epsilon_{\gamma }\gtrsim 80\%$, $50\% \lesssim \epsilon_{\gamma }\lesssim 80\%$, and $\epsilon _{\gamma }\lesssim 50\%$), the $E_{%
\text{p}}$ - $E_{\text{iso}}$ distribution \citep{Amati2002} and its dispersion can be ideally explained by the photosphere emission model. For the $\epsilon_{\gamma }\gtrsim 80\%$ sub-sample, the $E_{\text{p}}\propto (E_{\text{iso}})^{1/4}$ relation, the power-law shape of the X-ray afterglow and the significant reverse shock signals in the optical afterglow are revealed, supporting the photosphere model. For the $\epsilon_{\gamma}\gtrsim 50\%$ (including $\epsilon_{\gamma }\gtrsim 80\%$) sub-sample, a correlation of $\epsilon _{\gamma}=E_{\text{iso}}/E_{\text{k}} \simeq \eta /\Gamma$ is found, consistent with the prediction of the photosphere model. Here, $E_{\text{k}}$ is the remaining kinetic energy in the afterglow, $\eta$ means the baryon loading, and $\Gamma$ means the Lorentz factor of the jet.
For the $\epsilon_{\gamma}\lesssim 50\%$ sub-sample, a correlation of $ E_{\text{iso}}/E_{\text{k}} \simeq (R_{\text{ph}}/R_{s})^{-2/3}  \simeq E_{\text{ratio}}$ is obtained, coincided also with the photosphere model. Here, $R_{s}$ is the saturated acceleration radius, $E_{\text{ratio}}$ means the common decreased factor for $E_{\text{iso}}$ and $E_{\text{p}}$, deviating from the $E_{\text{p}}\propto (E_{\text{iso}})^{1/4}$ relation, due to adiabatic expansion. 
Fifth, using the theoretical photosphere spectrum (rather than the empirical function, BAND or CPL) to directly fit the observational data, excellent fitting results have been achieved both for the time-integrated spectrum and spectral evolutions \citep{Ahlgren2015,Ryde2017,Meng2018,Meng2019,Meng2024,Samuel2023}.

For the GRB emission from the nondissipative photosphere, in each local fluid element within certain angle ($\Delta \theta <~ 1/\Gamma$), scattering produces the
angular distribution of the emission, thus polarized \citep{Belo2011}. If
the jet is isotropic, when accounting for the emissions from the whole emitting region
simultaneously, the observed polarization signal can vanish due to the
rotational symmetry around the line-of-sight (LOS). However, if the jet is
structured and the viewing angle is nonzero, the symmetry should break and
certain polarization should exist. Several works have studied the dependence
of this photosphere polarization on jet structure \citep{Lund2014,Parso2020,Ito2024}, and found that the polarization degree is relatively low (a few $\%$) for typical structure,
and can reach 20$\%-40\%$ in extreme cases. For the GRB polarization from synchrotron radiation (relatively large, see \citealt{Lyuti2003,Waxman2003,Burge2019}), researches have shown that jets with ordered magnetic fields could produce high polarization
degrees ranging between $20\%$ and $70\%$ (see \citealt{Toma2009,Deng2016,Gill2020,Lan2021}), while jets with random magnetic fields produce smaller polarization degrees \citep{Lan2019,Tuo2024}. 

Motivated by the evidence for photosphere origin and the consistency of the photosphere polarization and the POLAR polarization, in this work, we analyze the characteristics of the prompt
emission, the X-ray afterglow, and the polarization degree for the GRBs with
reported polarization detection. Our findings support that these observations and their relationships can be explained well under the framework of the photosphere model. In particular, we discover the positive correlation between the low-energy spectral index $\alpha$ and the polarization degree.

Noteworthily, \citet{Li2025} have performed detailed spectral analysis for 26 bursts with significant polarization measurements. They find that the spectra of 10 bursts are best-fitted by the combination of a Band function and a blackbody, indicating a hybrid outflow. The usual existence (10/26) of the blackbody component implies that the photosphere emission can also be a possible mechanism for the high polarization. Similar to our work, the authors have also investigated how the polarization correlates to several key GRB properties, such as the $\alpha$. However, no robust correlation for $\alpha$ and PD is identified. The two main reasons for the different result from our work are as follows. First, they plot the distribution using the complete sample from POLAR, AstroSAT, and GAP. However, we consider the distributions for POLAR and AstroSAT, respectively. As stated in \citet{Chat2022}, the reported polarization degrees from AstroSAT are much higher than those from POLAR. All polarization degrees from AstroSAT are higher than $40\%$ (46$\%-94\%$), while most of the polarization degrees from POLAR are $\lesssim$ $20\%$ except for two bursts ($40\%$ and $60\%$; \citealt{Kole2020}). The PD discrepancy between POLAR and AstroSAT (along with GAP) will surely produce different regions in the PD-$\alpha$ distribution. Second, in \citet{Li2025}, for the bursts best-fitted by a Band function and a blackbody, the $\alpha$ is significantly affected by the additional blackbody component, being much softer (see the Table 2 therein, comparing with the Table 3 and Table 4; much $\alpha$ is $\lesssim$ -1). Almost all the $\alpha$ in Table 3 (best-fitted by a Band function) are harder than -0.9, and 4/9 of them are even $\gtrsim$ -2/3. Besides, the $\alpha$ for 5/7 of the bursts in Table 4 (best-fitted by a cutoff power law) are $\gtrsim$ -2/3. This quite hard $\alpha$ for the polarization sample is consistent with our result. And the consideration of an additional blackbody component is not necessary in our work, since the complex spectral structure of Band plus blackbody is solely produced by the photospheric emission.  

Our paper is organized as follows.  In Section \ref{sec:character}, we show the X-ray afterglow and the prompt emission characteristics for the GRBs with reported polarization detection. The power-law shape for the X-ray afterglows and the $T_{90}\propto (L_{\text{iso}})^{-0.5}$ correlation for the prompt emission are revealed, consistent with thermal origin. Then, in Section \ref{sec:alpha}, we further find the quite hard low-energy spectral index $\alpha$ and the positive correlation of the $\alpha$ and the polarization degree. Besides, we interpret this positive correlation well with the photosphere model.
In Section \ref{sec:Conclu}, we discuss the photosphere polarization and give a 
brief summary.

\section{The characteristics of the GRBs with reported polarization detections}

\label{sec:character}

\subsection{Power-law shape for the X-ray afterglows}

\begin{figure}[tbp]
\centering\includegraphics[angle=0,height=2.72in]{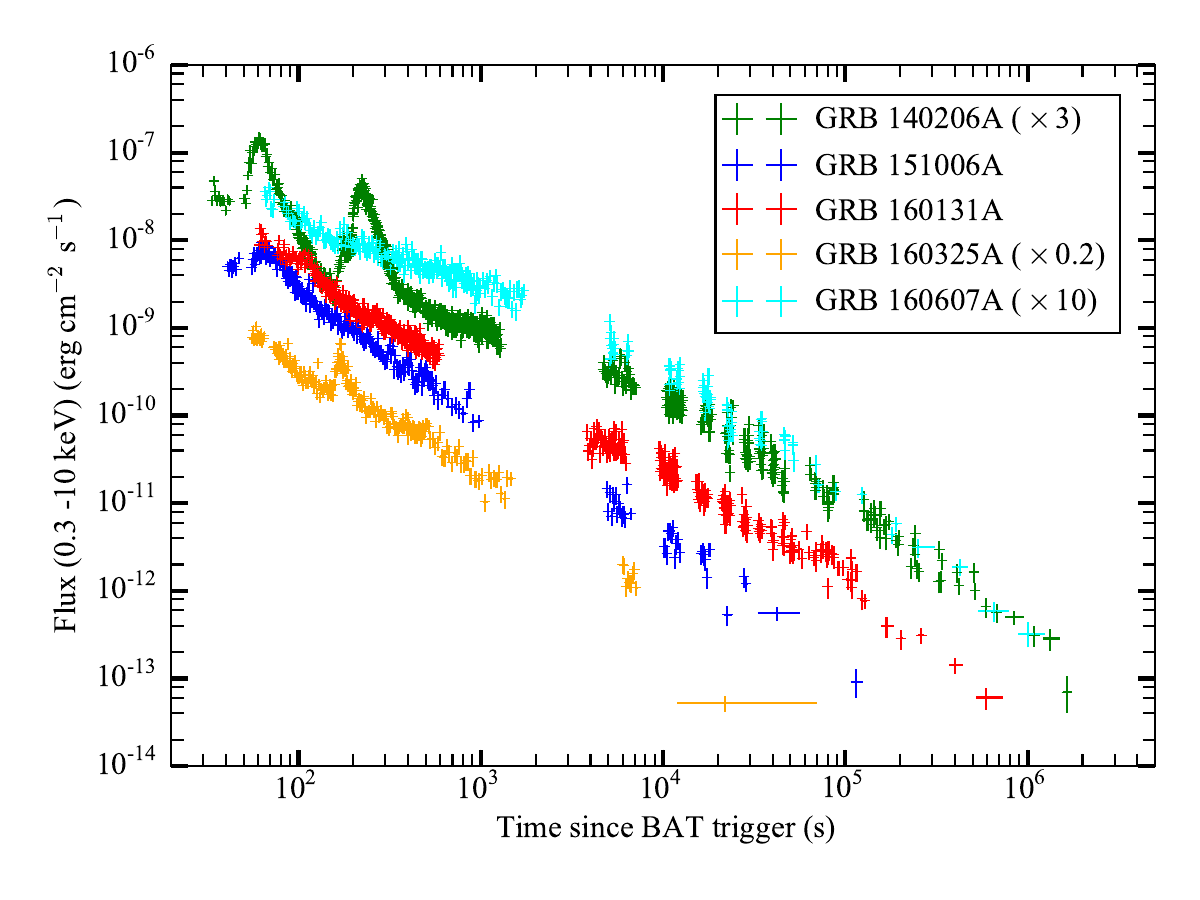} 
\vspace{-23pt}
\caption{The roughly power-law shape (without a significant plateau) for the X-ray afterglow light curves of the GRBs with reported polarization detections. This fits well with the prediction of the hot fireball model. The slight deviation from the power law (breaks) can arise from the moderate viewing angle $\theta_{v} \sim 1.5/\Gamma$ for the polarization sample (see \citealt{Rossi2002}). The early flares in GRB 140206A are inherited from the prompt emission, considering a very close match for the light curves of Swift/BAT and Swift/XRT, during this period.}
\label{fig:afterglow}
\end{figure}

\begin{figure}[tbp]
\centering\includegraphics[angle=0,height=2.58in]{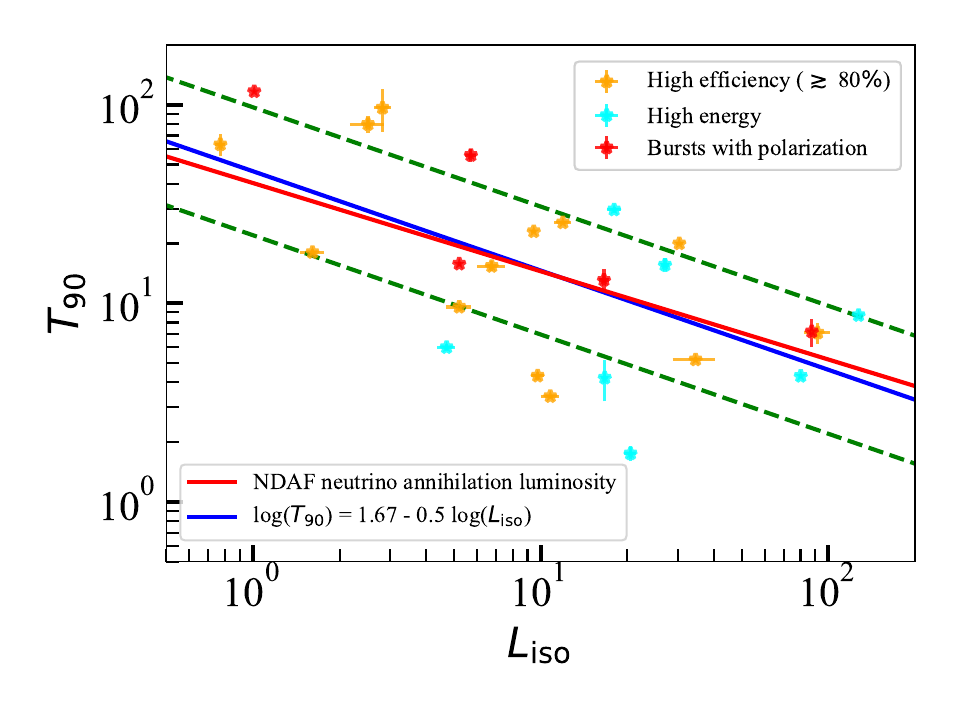} 
\caption{The $T_{90}\propto (L_{\text{iso}})^{-0.5}$ correlation (blue solid line) for the GRBs with reported polarization detections (red stars), along with the high-efficiency sample (orange stars; \citealt{Meng2022}) and the high-energy sample (cyan stars; $E_{\gamma,\text{beam}}$ $\gtrsim 10^{52}$ erg and obtaining GeV/TeV detection; \citealt{Sharma2021}). Notably, all three samples exhibit a roughly power-law shape in their X-ray afterglows. The dashed green lines mark a $T_{90}$ deviation of a factor of two, from the best-fit log ($T_{\text{90}}$) = 1.67 - 0.5 log ($L_{\text{iso}}$) relationship. This $T_{90}\propto (L_{\text{iso}})^{-0.5}$
correlation aligns closely with the predictions made by the NDAF model (red solid line), which is under the hot fireball framework.}
\label{fig:T90}
\end{figure}

In Figure \ref{fig:afterglow}, we show the X-ray afterglow light curves of the GRBs with reported polarization detections (including the upper and lower limits,
mainly from AstroSAT, see Table 1 in \citealt{Chat2019}). It is found that all the X-ray
afterglow light curves almost appear to be a simple power law, with a lack of
any plateau, steep decay, or significant flare (with weak flare in the early
time). This is quite similar to the results for the high-efficiency sample
in \citet{Meng2022}, the high-energy sample in \citet{Sharma2021} (with the jet-opening angle-corrected energy $E_{\gamma,\text{beam}}$ $\gtrsim 10^{52}$ erg, all obtaining the GeV/TeV detection) and the GeV/TeV-detected bursts in \citet{Yama2020}. The reason is that brighter bursts are selected to obtain the high statistical significance of the polarization measurement. In other words, it is also the sample with higher energy ($E_{\text{iso}}$ or $L_{\text{iso}}$). The millisecond magnetar has a maximum rotational energy of $\sim 10^{52}$ erg to form a jet \citep{Sharma2021}, and the power-law shape of the X-ray afterglow is the basic prediction of the classical hot fireball model for GRBs \citep{Paczy1993,Me1997}. Thus, the central engine for the GRBs with significant polarization is more likely to be a black hole.

\subsection{$T_{90}\propto (L_{\text{iso}})^{-0.5}$ correlation for the prompt emission}

In Figure \ref{fig:T90}, we show the $T_{90}$ (intrinsic) and $L_{\text{iso}}$
(adopted from \citealt{Xue2019}) distribution for the GRBs with reported polarization detections, along with
the high-efficiency sample and the high-energy sample. The sample with polarization and redshift detections (selected from Table 1 in \citealt{Gill2021}) consists of GRB 140206A, GRB 160131A, GRB 160509A, GRB 160623A and GRB 171010A, and we only consider the main bright pulse when obtaining the $T_{90}$ of GRB 120624B, GRB 160625B, GRB 190114C in the high-energy sample. Obviously, the $%
T_{90}\propto (L_{\text{iso}})^{-0.5}$ correlation, found in \citet{Meng2022}, is
available for all three samples (best-fitted by log ($T_{\text{90}}$) = 1.67 - 0.5 log ($L_{\text{iso}}$) for a combined sample). In the following, we stress that this
correlation is quite consistent with the prediction of the neutrino
annihilation model (for neutrino-dominated accretion flow, namely NDAF; see \citealt{Poph1999,Liu2017}), and
thus favors the thermal-dominated jet and the photosphere model.  The jet
power of neutrino annihilation can be approximated as (for dimensionless
black hole spin $a = 0.95$, see \citealt{Zala2011,Leng2014}):
\begin{eqnarray}
&& P_{\nu \overline{\nu}} \approx 1.3 \times 10^{52} \ \Big(\dfrac{M_{\text{BH}}}{3M_{\odot}}\Big)^{-3/2} \notag  \\
&& \times \left\{
\begin{array}{l}
\Big(\dfrac{\dot{M}}{M_{\odot}s^{-1}}\Big)^{9/4}, \ \ \ \dot{M}_{\text{ign}}<\dot{M}<\dot{M}_{\text{trap}}  \vspace{1ex} \\ 
\Big(\dfrac{\dot{M}_{\text{trap}}}{M_{\odot}s^{-1}}\Big)^{9/4}, \ \ \ \dot{M} \geq \dot{M}_{\text{trap}} 
\end{array}%
\right.  \     \text{erg} \ \text{s}^{-1},\label{b}
\end{eqnarray}
here $\dot{M}_{\text{ign}}=0.021 M_{\odot}s^{-1} (\alpha/0.1)^{5/3}$ is the ‘ignition’ accretion rate, $\dot{M}_{\text{trap}}=1.8 M_{\odot}s^{-1}(\alpha/0.1)^{1/3}$ is the ‘trapped’ accretion rate, and the $\alpha$ means the viscosity parameter of the neutrino-cooled disc. Note that the $P_{\nu \overline{\nu}}$ also depends on the dimensionless black hole spin $a$, an analytical correlation of $P_{\nu \overline{\nu}} \propto {[r_{\text{ms}} (a)/r_{g}]}^{-4.8}$ was obtained in \citet{Zala2011}, based on the numerical simulations. Here, the $a$-dependent $r_{\text{ms}}$ means the radius of the innermost stable orbit, $r_{g}=2GM/c^2$, and $r_{\text{ms}} (a)/r_{g}$ ranges from 3 ($a=0$) to 0.97 ($a = 0.95$). Additionally, other correlations have been proposed, such as $P_{\nu \overline{\nu}} \propto (10)^{2.45a}$ (see \citealt{Xue2013}).

As stated in \citet{Leng2014}, the average accretion rate $\dot{M}$ during a burst can be approximated as $\frac{M_{\text{BH}}}{T_{90}}$, assuming that the mass of the black hole roughly doubles through the accretion process. Also, it is considered that $L_{\text{iso}}=(4\pi/2\pi\theta_{\text{jet}}^2) \ \epsilon \ P_{\nu  \overline{\nu}}=(2\epsilon/\theta_{\text{jet}}^2)  P_{\nu  \overline{\nu}}$, here $L_{\text{iso}}$ is the observed isotropic equivalent luminosity from the jet, $\epsilon$ means the radiative efficiency, and $\theta_{\text{jet}}$ stands for the jet opening angle. Thus, the model-predicted $L_{\text{iso}}$ in the $\dot{M}_{\text{ign}}<\dot{M}<\dot{M}_{\text{trap}}$ regime for the long GRBs is (see also Equation (9) in \citealt{Leng2014}):
\begin{eqnarray}
L_{\text{iso}}=5.1 \times 10^{52} \dfrac{\epsilon}{0.3} \Big(\dfrac{\theta_{\text{jet}}}{0.1}\Big)^{-2} \Big(\dfrac{M_{\text{BH}}}{3M_{\odot}}\Big)^{3/4} \Big(\dfrac{T_{90}}{10\text{s}}\Big)^{-9/4} \ \text{erg} \ \text{s}^{-1},  \label{c}
\end{eqnarray}
where the reference values of $\epsilon=0.3$, $\theta_{\text{jet}}=0.1$ and $M_{\text{BH}}=3M_{\odot}$ are adopted. In the $\dot{M}>\dot{M}_{\text{trap}}$ regime, the optical depth of the neutrinos increases dramatically, then the neutrinos are trapped in the accretion disk and advected onto the black hole \citep{Di2002}. So, the $P_{\nu \overline{\nu}}$ and $L_{\text{iso}}$ rise much slower or approximate to be constant (no $T_{90}$ dependence is obtained; see Equation (\ref{b})). 

According to above, the neutrino annihilation model predicts a $T_{90}\propto (L_{\text{%
iso}})^{-0.44}$ correlation. Taken $\epsilon \equiv$ $0.9$ for the
high-efficiency sample, this model-predicted correlation (the red line
in Figure \ref{fig:T90}, multiplied by 1.5) is well consistent with the above observation. Interestingly, in Figure \ref{fig:T90}, for a few bursts with the smallest $T_{90}$
(less than 4 s), the $T_{90}$ drops much faster, close to $T_{90}\propto (L_{\text{iso}})^{-1}$. This can be naturally explained by the $\dot{M}$ $\gtrsim$ $\dot{M}_{\text{trap}}$ regime. On the one hand, the average accretion rate
$\dot{M}$ for these bursts should be larger than 3$M_{\odot}$/(4s) (close to $\dot{M}_{\text{trap}}$). On the other hand, both the numerical simulations in \citet{Zala2011} (see Figure 4 therein; our adopted Equation (\ref{b}) is a good analytical approximation derived from a simplified model) and in \citet{Lei2017} show that the luminosity (and $P_{\nu \overline{\nu}}$) in this regime should not remain completely constant. According to the solid lines and points in Figure 1 of \citet{Lei2017}, it is likely to exhibit a gradual rise as $\sim$ $L_{\text{iso}} \propto (\dot{M})^{1.0} \propto (M_{\text{BH}}/T_{90})^{1.0}$. Therefore, $T_{90}\propto (L_{\text{iso}})^{-1}$ is obtained.

Noteworthily, the above analysis favors the quite high dimensionless black
hole spin $a = 0.95$. In the next section, we find that the typical jet opening angle 
$\theta_{\text{jet}}$ for these bursts with higher energy could be two or three times
smaller than 0.1 ($\theta_{\text{jet}}$ $\sim$ 0.03 - 0.05). Then, the spin
may be smaller. If we take $L_{\text{iso}}\propto (10)^{2.45a}$ \citep{Xue2013}, the spin can be $\sim$ 0.54 - 0.7. Anyway, the black hole spin is
rather high, and the explicit value needs further exploration.

\section{The apparently positive correlation of the low-energy spectral index $\protect\alpha$ and the polarization degree}

\label{sec:alpha}

\subsection{Observational results}

\begin{figure}[tbp]
\centering\includegraphics[angle=0,height=2.7in]{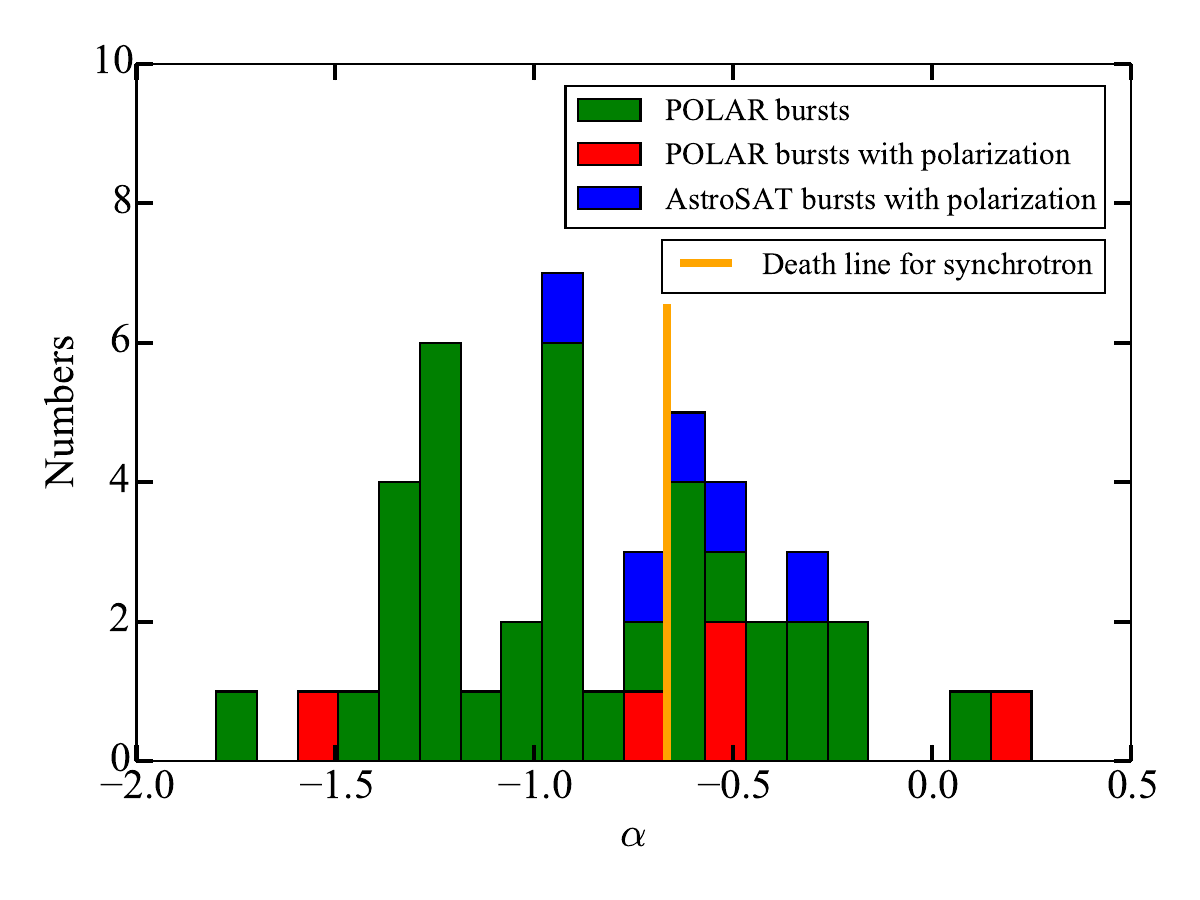} 
\vspace{-20pt}
\caption{The distribution of the low-energy spectral index $\alpha $, for the GRBs with reported polarization degree (the red and blue boxes, excluding the upper and lower limits). The $\alpha $
is quite close to or larger than the “synchrotron line of death” $\alpha $= $-$2/3 (the orange line). Note that GRB 170101A is only detected by POLAR
and \textit{Swift}/BAT, thus the $\alpha \equiv -1.55$ is doubtful.}
\label{fig:low_distri}
\end{figure}

\begin{table}
\caption{The correlation between $\alpha$ and the polarization degree.}
\label{TABLE:PD}\center%
\begin{tabular}{lcccl}
\toprule GRB Name & $\alpha$ &  Polarization degree \\
\midrule 
\textbf{POLAR:} &  \\
GRB 161218A &  -0.54 $\pm$ 0.07  &  9.0$_{-7.0}^{+10.7} \%$  \\
GRB 170114A &  -0.68  $\pm$ 0.09   &   4.0$_{-4.0}^{+10.5} \%$ \\
GRB 170127C &  0.25   $\pm$  0.12  &   11.0$_{-8.4}^{+19.3} \%$ \\
GRB 170206A &  -0.49  $\pm$  0.04  &   10.0$_{-8.6}^{+7.4} \%$ \\
\midrule
\textbf{AstroSAT:} &  \\
GRB 160106A & -0.53 $\pm$ 0.07 &  69 $\pm$ 24 $\%$ \\   
GRB 160325A  & -0.71 $\pm$ 0.07 &   59 $\pm$ 28  $\%$ \\
GRB 160802A  &  -0.61 $\pm$ 0.04 &  85 $\pm$ 30 $\%$ \\
GRB 160821A & -0.97 $\pm$ 0.01 &  54 $\pm$ 16 $\%$ \\
GRB 160910A & -0.36 $\pm$ 0.03 &  94 $\pm$ 32 $\%$ \\
\midrule
\end{tabular}
\tablefoot{
We show the distribution of the low-energy spectral index $\alpha$ and the polarization degree for 4 POLAR bursts (adopted from \citealt{Zhang2019}) and 5 AstroSAT bursts (adopted from \citealt{Chat2019}) . We do not adopt GRB 170101A in POLAR and
160131A in AstroSAT due to lack of \textit{Fermi}/GBM detection. The low-energy spectral index $\alpha$ from \textit{Swift}/BAT or Konus/Wind could have relatively large error. The errors of PD for POLAR are taken from \citet{Kole2020}.
}
\end{table}

\begin{figure*}[tbp]
\centering\includegraphics[angle=0,height=2.3in]{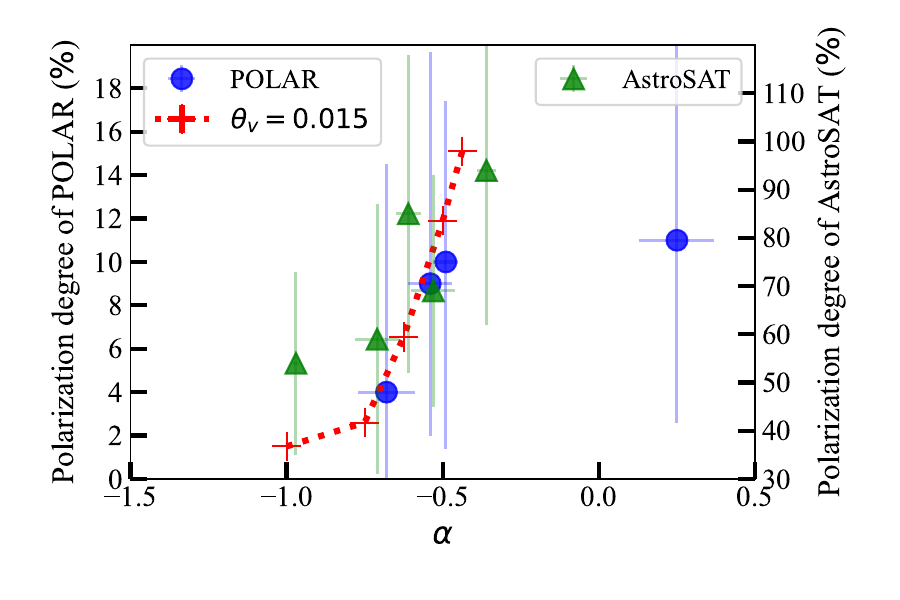} 
\put(-3.65in,2.0in){\bf (a)}
\centering\includegraphics[angle=0,height=2.26in]{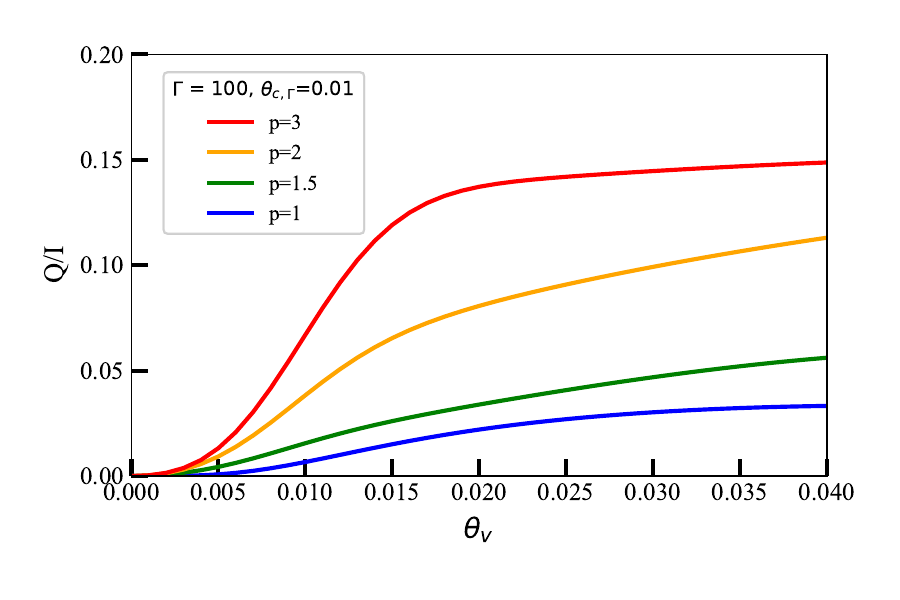} 
\put(-3.59in,2.0in){\bf (b)}
\vspace{-7pt}
\caption{The apparently positive correlation of the low-energy spectral index $\alpha$ and the polarization degree (left) and the possible photosphere explanation (right, polarization degree $\Pi =\left\vert Q\right\vert /I$). (a) Notably, this positive correlation can be obtained from the POLAR sample (blue circles, with Pearson correlation coefficient r=0.69) and the AstroSAT bursts (green triangles, r=0.85), both. The P values standing for the significance level are both $\sim$ 0.06, thus the correlation is almost significant (combining POLAR and AstroSAT, P $\sim$ 0.03 can be achieved). (b) For the photosphere polarization, the polarization degree is shown to be positively correlated with the p value (the power-law decreasing index of the $\Gamma$, for the jet angular distribution), for the smaller $\theta_{c,\Gamma}$ case available for bursts with higher energy. Also, for $\Gamma \cdot \theta_{c,\Gamma} \simeq 1$, $\alpha$ is positively correlated with the p value, $\alpha \simeq (-1/4) (1+3/p)$. The predicted positive correlation of $\alpha$ and the polarization degree for $\theta_v$=0.015 (red pluses in the left panel) can match the observations of POLAR and the slope of the AstroSAT sample, approximately. Notice that the GRB 170127C in POLAR may have larger $\theta_{c,\Gamma}$, thus possessing larger $\alpha$ of 0.25 and smaller polarization degree (see Figure \ref{fig:theta_c} (a), maybe two to three times smaller).}
\label{fig:alpha_PD}
\end{figure*}

\begin{figure*}[tbp]
\centering\includegraphics[angle=0,height=2.47in]{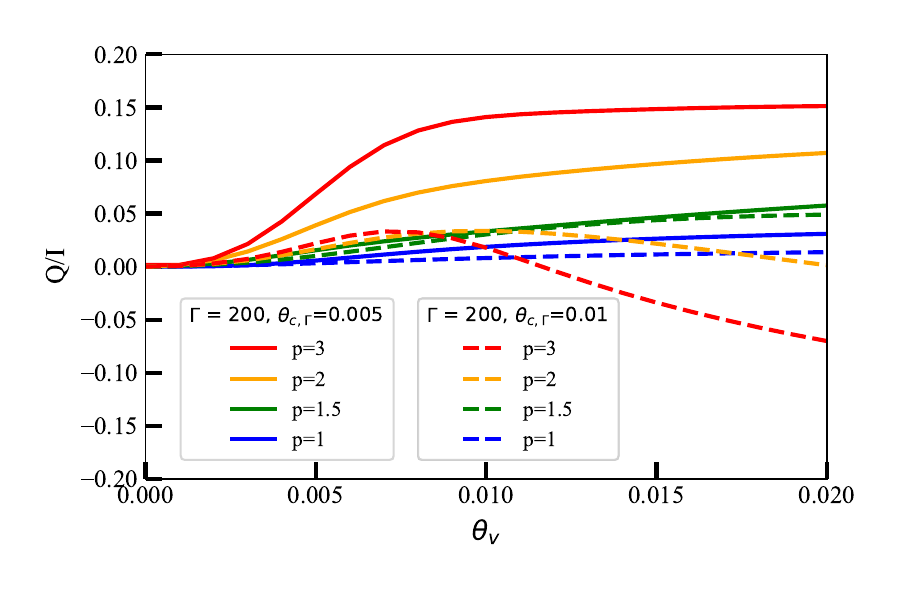} 
\put(-3.83in,2.2in){\bf (a)}  
\centering\includegraphics[angle=0,height=2.48in]{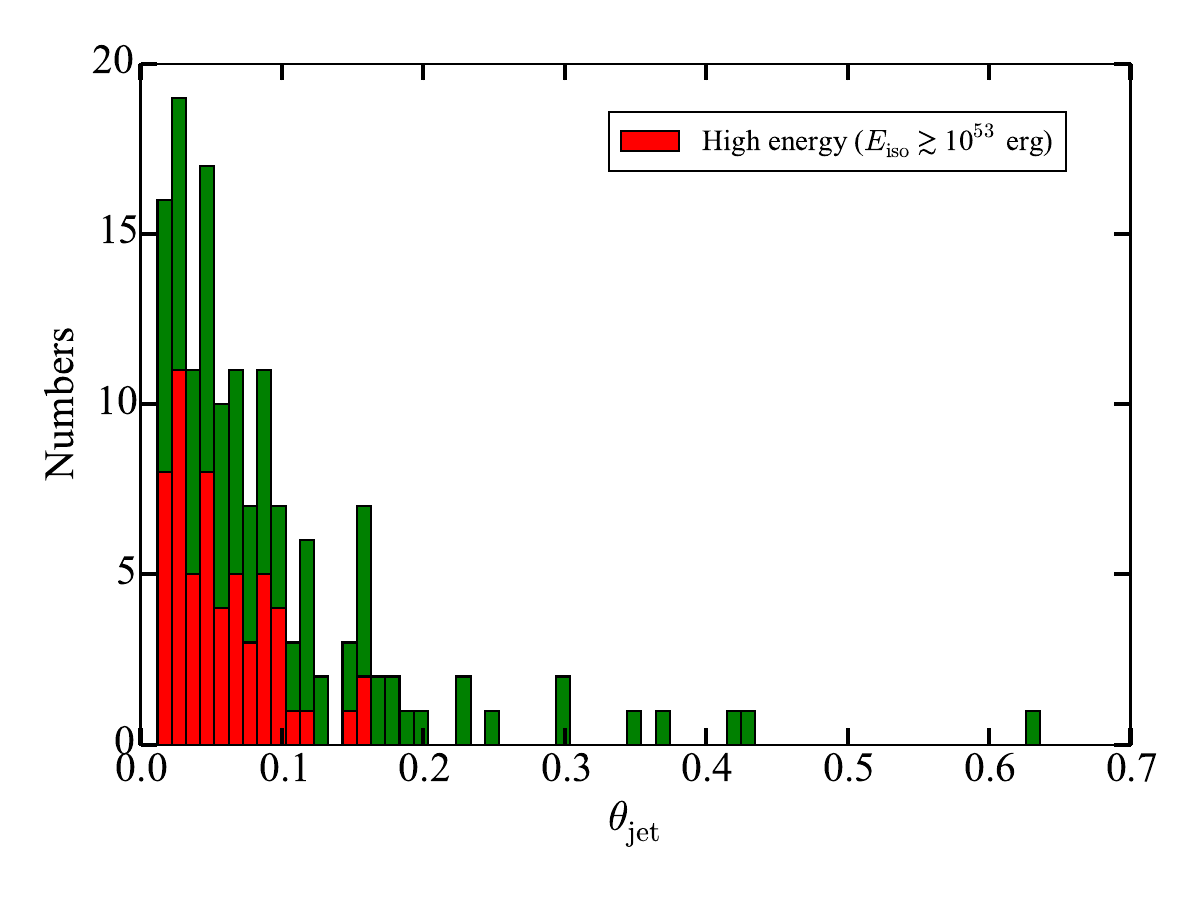} 
\put(-3.4in,2.2in){\bf (b)}
\vspace{-5pt}
\caption{Correlations of photosphere polarization degree ($\Pi =\left\vert Q\right\vert /I$), $\theta_{c,\Gamma}$ (or $\theta _{\text{jet}}$) and $E_{\text{iso}}$. (a) Comparison of the photosphere polarization degree for smaller $\theta_{c,\Gamma}$ ($\Gamma \cdot \theta_{c,\Gamma}=1$, solid lines; predicting much larger polarization degree) and larger $\theta_{c,\Gamma}$ ($\Gamma \cdot \theta_{c,\Gamma}=2$, dashed lines; predicting much smaller polarization degree). (b) The smaller $\theta _{\text{jet}}$ (likely $\theta_{c,\Gamma}$ also) is found for the high-energy sample ($E_{\text{iso}} \gtrsim 10^{53}$ erg), indicating the polarization detection.}
\label{fig:theta_c}
\end{figure*}

In Figure \ref{fig:low_distri} (see also Table \ref{TABLE:PD}), we show the distribution of the low-energy spectral index $\alpha $, for the GRBs with reported polarization degree
(omitting the upper and lower limits). This sample consists of the 5 bursts
reported by POLAR finally (with incoming angle below 45$^\circ$, see \citealt{Zhang2019}),
and 5 bursts reported by AstroSAT (5 upper limits are excluded, and GRB
160131A is excluded due to lack of \textit{Fermi}/GBM detection, see \citealt{Chat2019}). The
low-energy spectral index is taken from Table 1 in \citet{Kole2020} for POLAR, and Table
1 in \citet{Chat2019} for AstroSAT. Obviously, the low-energy spectral index $\alpha $ for most of these bursts (except for GRB 170101A in POLAR and GRB 160821A in AstroSAT) is rather hard, quite close to or larger than the "synchrotron line of death" ($\alpha $= $-$2/3; \citealt{Preece1998}). This terminology means that $\alpha$ should not exceed the value of -2/3 corresponding to the slow cooling limit, for the synchrotron shock model. When considering much faster cooling of particles, $\alpha$ should be much softer, extending to a limit of -3/2. Thus, the polarization for these bursts is more likely to be produced by the photosphere emission. Note
that, we do not conclude that the polarization for all GRBs comes from the
photosphere emission. But, for the polarization sample so far and in the
near future, which has higher energy, the photosphere may be the more
favorable mechanism. Also, notice that GRB 170101A is only detected by POLAR
and \textit{Swift}/BAT, and the best-fit peak energy $E_{\text{p}}$ is quite large
(323 keV, close to or beyond the detection upper limit), meaning that the
best-fit model is actually the PL (power law) model. Thus, the low-energy
spectral index $\alpha \equiv -1.55$ is doubtful, or the origin of this burst may be special (with high intrinsic luminosity and extremely large viewing angle).

In Figure \ref{fig:alpha_PD} (a) (see also Table \ref{TABLE:PD}), we illustrate the distribution of the low-energy spectral index $\alpha $ and the polarization degree, for the above 4 POLAR bursts (GRB 170101A is excluded) and 5 AstroSAT bursts. Interestingly, an obviously positive correlation of $\alpha$ and PD exists, for both the POLAR sample and the AstroSAT sample. The Pearson correlation coefficient is r=0.69 for the POLAR sample, and r=0.85 for the AstroSAT sample. Both are positive and larger than 0.6, indicating a strong positive correlation. Besides, the P values representing the significance level are both $\sim$ 0.06, thus the correlation is nearly significant ($\sim$ 5$\%$). When combining the data of POLAR and AstroSAT (accounting for their PD difference), P $\sim$ 0.03 ($\le$ 5$\%$; then significant) can be achieved. It is worth mentioning that the higher polarization degree for AstroSAT may be due to its higher energy band and narrower bandwidth, considering the photosphere emission model (see Section \ref{sec:difference}). As stated in Section \ref{sec:Intro}, \citet{Li2025} obtained the nondependence of $\alpha$ and PD, ignoring the PD difference between POLAR and AstroSAT. Meanwhile, we consider a smaller and more reliable (likely) POLAR sample from \citet{Zhang2019}, with additional criteria of higher fluence and smaller incident angle of $\leq$ 45$^{\circ }$. In Section \ref{sec:POLAR}, we again achieve the positive correlation between $\alpha$ and PD, using a larger POLAR sample from \citet{Kole2020} alone.    

To further test the significance of the $\alpha$-PD correlation, we perform the regression analysis and account for the uncertainties of the fitted parameters following \citet{Kelly2007}. For the combined data of POLAR and AstroSAT (accounting for their PD difference by taking a decreased factor for AstroSAT), we obtain $PD=(10.79 \pm 1.13)+(4.95 \pm 1.87) \cdot \alpha$, thus the $\alpha$ dependence of PD is significant at 2.65 $\sigma$. For the POLAR data alone, $PD=(26.18 \pm 1.88)+(32.47 \pm 3.27) \cdot \alpha$ (omitting the outlier of GRB 170127C; $10 \ \sigma$) and $PD=(10.37 \pm 1.96)+(5.12 \pm 3.82) \cdot \alpha$ (including GRB 170127C; $1.34 \ \sigma$) are achieved. For a larger POLAR sample (11 bursts) from \citet{Kole2020}, $1.72 \ \sigma$ is acquired (see Section \ref{sec:POLAR}). For the AstroSAT data alone, $PD=(112.71 \pm 15.28)+(63.70 \pm 22.89) \cdot \alpha$ (significant at 2.78 $\sigma$) is obtained. Besides, when adding more AstroSat bursts from \citet{Chat2022} as stated in the next paragraph, the combined data of POLAR and AstroSAT can derive 3.05 $\sigma$. According to the above results ($\gtrsim$ 2 $\sigma$), we can claim that the positive correlation between $\alpha$ and PD is almost significant.

Notice that, for the five-year AstroSat polarization sample in \citet{Chat2022}, the correlation of $\alpha $ and the polarization degree is also analyzed (see Figure 5(b) therein; for GRB 180103A, GRB 180120A, GRB 180427A, GRB 180914B, and GRB 190530A). They claim that no significant trend is observed. However, the apparent positive correlation exists without considering the GRB 180103A (maybe GRB 180120A also). By carefully checking the properties of GRB 180103A, we think the inclusion is improper because of the following reasons: (1) the light curve of this burst appears as the complex multi-pulse (three or four pulses, with different shapes and amplitudes) structure, with extremely long duration of $T_{90} \sim 165.83$ s. The soft $\alpha= -1.31$ is for the overlapping of different pulses, for the single bright pulse (measured from $T_{0}+81.664$ s to $T_{0}+93.952$ s) $\alpha= -1.00$ (see GCN Circular 22314). (2) This burst is only detected by \textit{Swift}/BAT and Konus/Wind, and lack of \textit{Fermi}/GBM detection. Due to the relatively high low-energy cut of Konus/Wind (20 keV - 20 MeV) and the low $E_{p} \sim 273$ keV of this burst, the measured $\alpha= -1.00$ is likely to suffer from the exponential component and thus mistakenly softer (The $\alpha$ obtained from Konus/Wind seems to be softer than that from \textit{Fermi}/GBM, see Figure 2 (b) in \citealt{Meng2024}). (3) The incident angle of this burst is quite large $\sim$ 52.33$^{\circ }$ ($\gtrsim$ 45$^{\circ }$), so the polarization is greatly influenced by the interactions with the satellite elements. Besides, the light curve of GRB 180120A consists of two comparable peaks, and the error of the polarization degree is the largest among five bursts, 62.37 $\pm$ 29.79 $\%$.

The extremely hard $\alpha= -0.29$ of GRB 180427A, significantly 
exceeding the “synchrotron line of death” and obtaining from \textit{Fermi}/GBM, strongly supports its photosphere origin. Meanwhile, its polarization degree of 60.01 $\pm$ 22.32 $\%$ is extraordinarily large. These two properties are well in line with GRB 160910A ($\alpha= -0.36$) in the AstroSAT polarization sample, and GRB 170127C ($\alpha= 0.25$) in the POLAR polarization sample (see Table \ref{TABLE:PD}). In addition, for the polarization sample of GAP (on board the IKAROS solar power sail; GRB 100826A, GRB 110301A, and GRB 110721A; \citealt{Yone2011,Yone2012}), GRB 110721A is the well-known burst with thermal component \citep{Axel2012}. And, the $\alpha$ for GRB 100826A ($\alpha= -0.81$ for \textit{Fermi}/GBM, taken from Table 1 in \citealt{Guan2023}; $\alpha= -0.84$ for Konus-Wind, see GCN 11158) and GRB 110301A ($\alpha= -0.81$ for \textit{Fermi}/GBM, see GCN 11771) are both quite hard. Notice that several studies \citep[e.g.,][]{Burgess2020} have shown that bursts with steep $\alpha$ can still be explained by synchrotron radiation.

\subsection{Theoretical explanation}

For synchrotron emission, with a locally ordered magnetic field (maximum
linear polarization is obtained), the local polarization degree depends on the low-energy spectral (photon) index\footnote{Note that in several works involving the polarization of synchrotron emission \citep{Gill2020,Guan2023}, the low-energy spectral index $\alpha_{\text{s}}$ stands for the flux spectrum, and $\alpha= - \alpha_{\text{s}} - 1$.} $\alpha$ as (see \citealt{Toma2013,Gill2020} and Equation (10) in \citealt{Gill2021}): 
\begin{eqnarray}
\Pi _{\text{max}}\equiv \frac{\alpha }{\alpha -2/3}.
\end{eqnarray}
Thus, the softer $\alpha $, the larger polarization is obtained. Namely, the
negative correlation of $\alpha $ and the polarization degree is predicted,
which is against the above observation. For other magnetic field structures
(with B$_{\bot }$ component), the polarization degree is a fraction of the above maximum polarization, depending on the polar angle. So, this negative
correlation should also be held.

For the nondissipative photospheric emission from a structured jet, the
global polarization degree $\Pi =\left\vert Q\right\vert /I$ can be
calculated from (see also Equations (11) in \citealt{Lund2014}):
\begin{eqnarray}
\frac{Q}{I}=\frac{\int_{\Omega _{\text{s}}}D^{2}(d\dot{N}/d\Omega
)\Pi (\theta _{\text{L}})\cos (2\phi _{\text{L}})d\Omega }{%
\int_{\Omega _{\text{s}}}D^{2}(d\dot{N}/d\Omega )d\Omega },  \label{c}
\end{eqnarray}%
where $\Omega$ ($\theta _{\text{L}},\phi _{\text{L}}$) is the angular coordinates of each local fluid element, relative to the LOS. $\phi _{\text{L}}=0$ means that the fluid element is in the plane of the LOS and the jet symmetric axis. $D=[\Gamma (1-\beta \cos \theta _{\text{L}})]^{-1}$ is the Doppler factor, and $D^{2}$ stands for the angular probability distribution of emitting photons. $d\dot{N}/d\Omega $ is the angular distribution of photon number (or luminosity) for the structured jet, with jet boundary angle of $\theta _{\text{s}}$ (solid angle is $\Omega _{\text{s}}$). In addition, $\Pi (\theta _{\text{L}})$ is the approximated polarization degree of each fluid element, taking the form of: 
\begin{eqnarray}
\Pi (\theta _{\text{L}})\simeq 0.45\frac{(1-\beta \cos \theta _{\text{L}%
})^{2}-(\cos \theta _{\text{L}}-\beta )^{2}}{(1-\beta \cos \theta _{\text{L}%
})^{2}+(\cos \theta _{\text{L}}-\beta )^{2}}.  \label{cd}
\end{eqnarray}
Here, the angular dependence comes from the Thompson scattering. A
factor of $0.45$ is taken because the emission moving at a comoving angle of $\pi/2$ is shown to have $\Pi \approx $ 0.45 (rather than 1.0), close
to and above the photosphere in a spherical outflow (see Figure 6 in \citealt{Belo2011}).

In this work, similar to previous works \citep{Meng2019,Meng2022b,Meng2024}, we consider the structured jet with the following assumptions: 
\begin{eqnarray}
\renewcommand{\arraystretch}{1.4}
\left\{ 
\begin{array}{l}
\Gamma (\theta ) =(\Gamma_{0}-1.2)/[(\theta /\theta _{c,\Gamma})^{2p}+1]^{1/2}+1.2, \\
L (\theta ) =L_{0}/[(\theta /\theta _{c,L})^{2q}+1]^{1/2}, \\
\theta _{c,\Gamma }<\theta _{c,L}, \  \  \theta _{v}<\theta _{c,L}.
\end{array}
\right.
\end{eqnarray}
Here, $\Gamma (\theta )$ represents the angular distribution for Lorentz factor, $\Gamma _{0}$ is the Lorentz factor in the center isotropic core, $\theta _{c,\Gamma}$ is the half opening angle of this $\Gamma$-constant core, and p denotes the power-law decreasing index of $\Gamma$ outside the core. Similarly, $L_{0}$ is the constant luminosity in the center core, and $\theta _{c,L}$ denotes the angular width for this core. Besides, $\theta _{c,\Gamma}<\theta _{c,L}$ is accounted for, based on the jet simulations \citep{Zhang2003,Geng2019} and the enhanced material density at a larger angle (see detailed discussions in \citealt{Meng2022b}). Normally, we should obtain $\theta _{v}<\theta _{c,L}$ to observe the bright bursts (the observed luminosity has already dropped when $\theta _{v} >\theta _{c,\Gamma}$, since the $\Gamma$ decreases). For this case, the injected luminosity distribution can be approximated as angle-independent since the photospheric emission is emitted from a narrow region of $\sim$ 5/$\Gamma$ (the jet boundary $\theta _{s}<=\theta _{c,L}$ is adopted). 

Based on Equation $(\ref{c})$, in Figure \ref{fig:alpha_PD}(b), with $\Gamma _{0}=100$ and $\theta _{c,\Gamma}=0.01$  ($\theta _{c,\Gamma}*\Gamma _{0}=1$, narrower core), we show the photosphere polarization degree for different $\theta _{v}$ and p. A similar result for p=4 can be seen in Figure 2 of \citet{Lund2014}. From Figure \ref{fig:alpha_PD}(b), it is obvious that the polarization degree strongly depends on the p value. With a larger p value, a higher polarization degree is obtained. 

Noteworthily, for the narrow core ($\theta _{c,\Gamma}*\Gamma _{0}<=3$), the low-energy spectral index $\alpha $ also strongly depends on the p value as (see also Equation (26) in \citealt{Lund2013}):
\begin{eqnarray}
\alpha \simeq (-1/4) (1+3/p),
\end{eqnarray}
and weakly depends on the $\theta _{v}$ (see Figure 6 in \citealt{Lund2013} and Figure 3 (e) in \citealt{Meng2019}). This equation tells that, with larger p value, the higher $\alpha$ is obtained. So, a positive correlation between the $\alpha$  and the polarization degree is predicted theoretically.

To further test this, in Figure \ref{fig:alpha_PD}(a), we plot the theoretical distribution of the $\alpha$  and the polarization degree for $\theta _{v}=$ 0.015 (see the other adopted  parameters and the parameter dependence for PD in Figure \ref{fig:alpha_PD}(b), $\Gamma \cdot \theta_{c,\Gamma}=1$). Obviously, the observed correlation of POLAR and the slope of the AstroSAT sample can be explained. The GRB 170127C in POLAR may have larger $\theta_{c,\Gamma}$, thus possessing larger $\alpha$ of 0.25 and smaller polarization degree (see Figure \ref{fig:theta_c} (a), maybe two to three times smaller).

In Figure \ref{fig:theta_c} (a), we compare the photosphere polarization degree for different $\theta_{c,\Gamma}$, $\Gamma \cdot \theta_{c,\Gamma}=1$ and $\Gamma \cdot \theta_{c,\Gamma}=2$. A much larger polarization degree is predicted for smaller $\theta_{c,\Gamma}$, with $\Gamma \cdot \theta_{c,\Gamma}=1$. Also, in Figure \ref{fig:theta_c} (b), we compare the observed $\theta_{\text{jet}}$ (taken from \citealt{Du2021}) for bursts with $E_{\text{iso}} \gtrsim 10^{53}$ erg and $E_{\text{iso}} \lesssim 10^{53}$ erg, respectively. The $\theta_{\text{jet}}$ for the high-energy sample ($E_{\text{iso}} \gtrsim 10^{53}$ erg) is shown to be significantly smaller, also implying the $\theta_{c,\Gamma}$. Note that this trend is well consistent with GRB 221009A (the BOAT burst, Brightest Of All Time), which is claimed to possess an extremely narrow jet ($\sim$ 1 $^{\circ }$, or 0.017 rad, see \citealt{Cao2023}). Thus, for the high-energy bursts, the large flux should enhance the polarization detection and the smaller $\theta_{c,\Gamma}$ can increase the intrinsic polarization degree, which make them the ideal candidates for polarization detection. Meantime, the millisecond magnetar has a maximum rotational energy of $\sim 10^{52}$ erg to form a jet \citep{Sharma2021}. Thus, for the bursts with higher energy and potential polarization detection, the central engine is more likely to be a black hole. Additionally, considering the probable power-law shape of the X-ray afterglow (see Figure \ref{fig:afterglow}), which matches the prediction of the hot fireball model, the photosphere origin of the prompt emission is preferred.

From Figure \ref{fig:alpha_PD}(b) and Figure \ref{fig:theta_c} (a), it is evident that the jet has to be observed substantially off-axis ($\theta _{v} > \theta_{c,\Gamma}$) to obtain significant polarization. This may imply an issue of explaining polarization with photospheric model that the off-axis observation will cause an important reduction of the GRB luminosity. Especially, when we try to explain the higher PD of AstroSAT (see Section \ref{sec:difference}) , the $\theta _{v} \gtrsim 4 \cdot \theta_{c,\Gamma}$ condition may be considered. For the POLAR sample, we can effectively interpret it with $\theta _{v} \sim 1.5 \ \theta_{c,\Gamma}$ (see Figure \ref{fig:alpha_PD}(a)). In this case, for $p=3$ (PD $\sim 12\%$), the observed $\Gamma$ falls down by $(1.5)^{p}=3.375$ times and the luminosity ($L \propto \Gamma^{8/3}$) decreases by a factor of $\sim$ 25.6. Thus, the issue is not particularly significant in this condition.

\subsection{$\alpha$ - PD distribution for a larger POLAR sample}

\label{sec:POLAR}

\begin{figure}[tbp]
\centering\includegraphics[angle=0,height=2.7in]{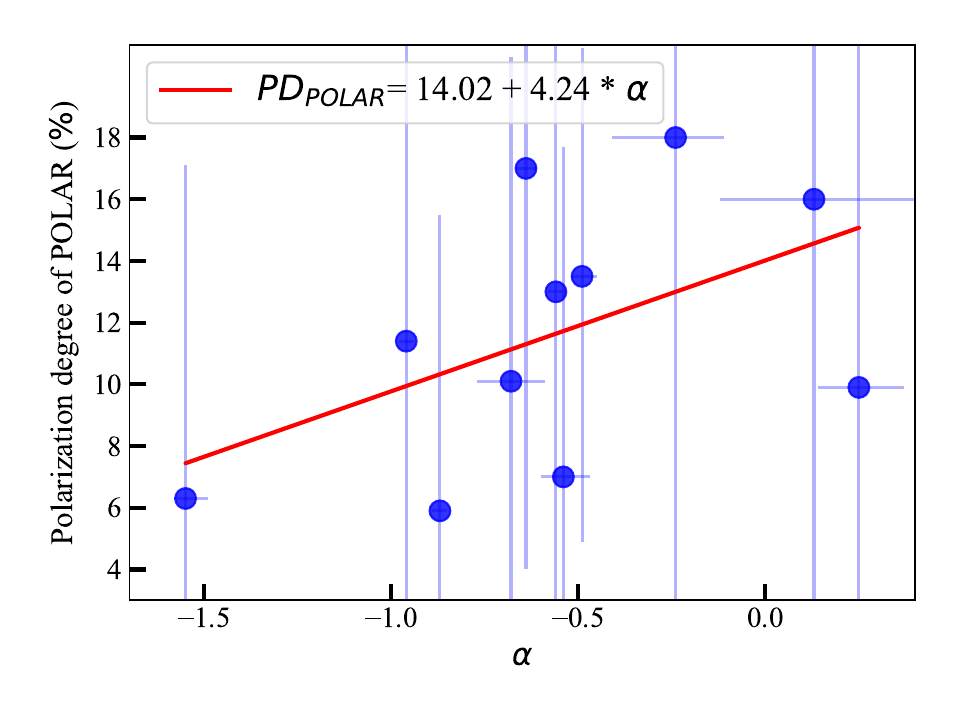} 
\vspace{-23pt}
\caption{The $\alpha$ - PD  distribution for a larger POLAR sample (blue circles, 11 bursts) in \citet{Kole2020}. It can be seen that the positive correlation between $\alpha$ and PD is quite obvious. The red solid line shows the best-fit correlation, which is $PD=(14.02 \pm 1.81)+(4.24 \pm 2.47) \cdot \alpha$. The correlation is almost significant, with a significance level of $1.72 \ \sigma$.}
\label{fig:polar}
\end{figure}

To further check the $\alpha$ - PD correlation, in Figure \ref{fig:polar}, we show the $\alpha$ - PD distribution for a larger POLAR sample in \citet{Kole2020}. We use 11 bursts with PD $<$ 20$^{\circ }$, omitting three obvious outliers (PD=60$^{+24}_{-36}$,  40$^{+25}_{-25}$ and 21$^{+30}_{-16}$, respectively). Another motivation for the selection is that the photosphere polarization is typically $<$ 20$^{\circ }$ (see Figure \ref{fig:alpha_PD}), unless the viewing angle is rather large ($\theta _{v} \gtrsim 3 \cdot \theta_{c,\Gamma}$). From Figure \ref{fig:polar}, the positive correlation between $\alpha$ and PD is apparently obtained, which is best fitted by $PD=(14.02 \pm 1.81)+(4.24 \pm 2.47) \cdot \alpha$. Thus, it is nearly significant at $1.72 \ \sigma$. Notice that an approximate slope of 4.37 is achieved for the complete sample of 14 bursts. Besides, the best-fit slope remains positive for arbitrary burst number, which is also available for the AstroSAT sample (combining the samples in \citealt{Chat2019} and \citealt{Chat2022}).

\subsection{Possible explanation for the higher PD of AstroSAT}

\label{sec:difference}

\begin{figure*}[tbp]
\centering\includegraphics[angle=0,height=2.4in]{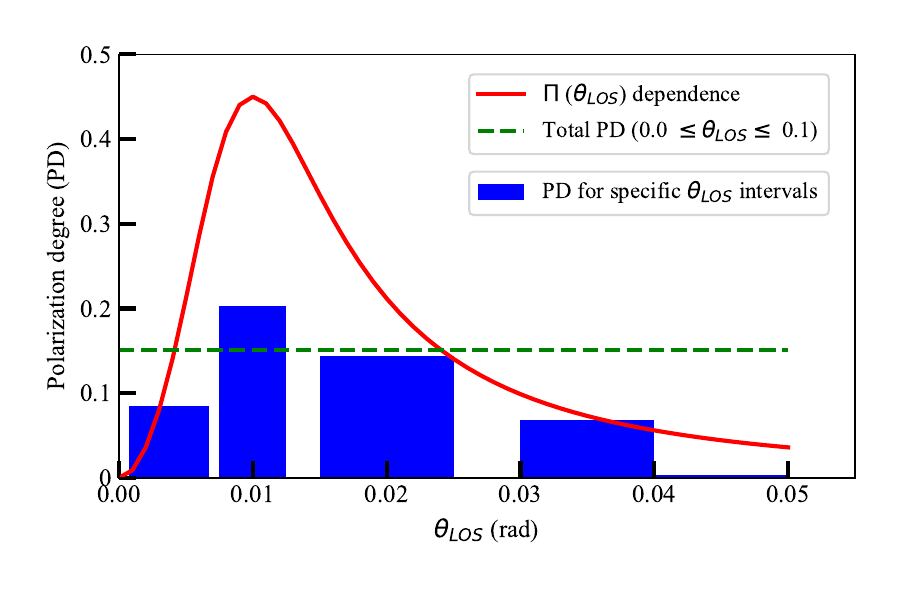} 
\put(-3.4in,2.0in){\bf (a)}  
\centering\includegraphics[angle=0,height=2.4in]{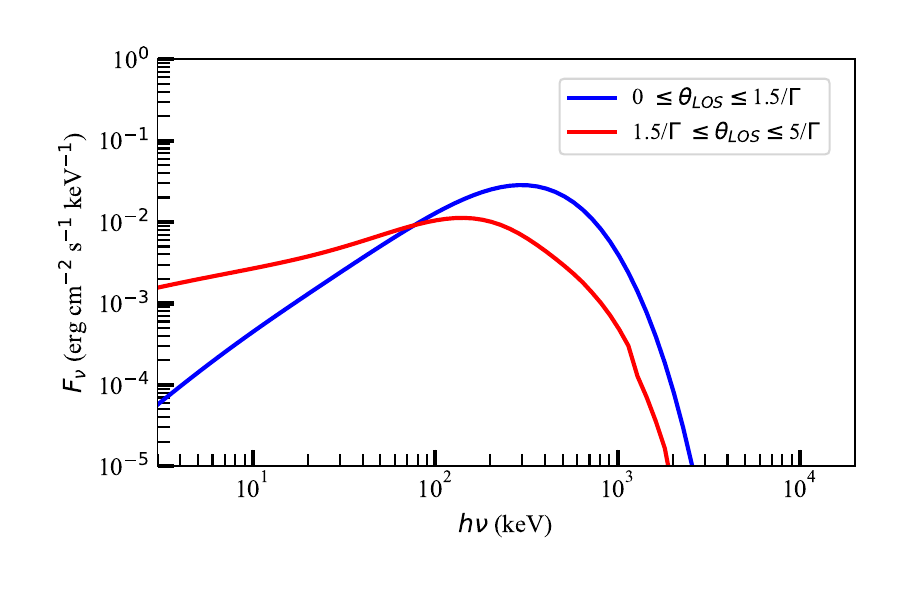} 
\put(-3.33in,2.0in){\bf (b)}
\vspace{-16pt}
\caption{A possible explanation for the PD discrepancy between POLAR and AstroSAT. (a) The PD distribution for the emission from different local fluid elements, relative to the light of sight. $\Gamma=100$, $\theta_{c,\Gamma}=0.01$, $\theta_{v}=0.015$, and p=4 are adopted. The red solid line represents the polarization degree of each fluid element with $\theta _{\text{LOS}}$, without considering the influence of azimuthal angle $\phi _{\text{LOS}}$ (see Equation (\ref{cd})). The blue boxes illustrate the integrated polarization degree for specific $\theta_{\text{LOS}}$ intervals, considering the influence of azimuthal angle. Obviously, the polarization degree rapidly rises, peaks at around $\theta _{\text{LOS}} \sim 1/\Gamma$, and then drops down quickly. The high polarization is almost within $\sim 1.5/\Gamma$ or $\sim 2/\Gamma$. The green dashed line marks the total polarization degree for $0 \leq \theta _{\text{LOS}} \leq 0.1$. (b) The comparison of the spectra emitted from $0 \leq \theta _{\text{LOS}} \leq 1.5/\Gamma$ and $1.5/\Gamma \leq \theta _{\text{LOS}} \leq 5/\Gamma$ ($\Gamma=200$, $\Gamma \cdot \theta_{c,\Gamma}=1$, and $\theta_{v}=1.5/\Gamma$ are adopted). Obviously, the spectrum from the inner region ($0 \leq \theta _{\text{LOS}} \leq 1.5/\Gamma$) has higher energy and higher PD (according to the left panel). This may contribute to the larger PD of AstroSAT (100 keV–600 keV), due to its higher band energy and narrower bandwidth than POLAR (50 keV–500 keV). The narrow bandwidth of $\sim$ 6 times will enable only the detection of the high-energy component from the inner region (the right portion of the blue line).}
\label{fig:higher_PD}
\end{figure*}

In Figure \ref{fig:higher_PD} (a), considering the photosphere emission from a structured jet, we show the PD distribution for the emission from different local fluid elements ($\theta_{\text{LOS}}$), relative to the light of sight. Simplistically, the red solid line calculates the polarization degree for each fluid element of $\theta_{\text{LOS}}$ according to Equation (\ref{cd}), without considering the influence of azimuthal angle $\phi_{\text{LOS}}$ (namely $\phi_{\text{LOS}}=0$; $\Gamma=100$ is adopted). More accurately, the blue boxes are the integrated results calculated from Equation (\ref{c}), accounting for the influence of azimuthal angle ($\phi_{\text{LOS}}$ is taken from 0 to $2*\pi$) and different specific $\theta_{\text{LOS}}$ intervals. $\Gamma=100$, $\theta_{c,\Gamma}=0.01$, $\theta_{v}=0.015$, and p=4 are adopted. It can be seen that the polarization degree rapidly increases, peaks at around $\theta _{\text{LOS}} \sim 1/\Gamma$, and then falls quickly. So, the high polarization is almost within $\theta_{\text{LOS}} \sim 1.5/\Gamma$ or $\sim 2/\Gamma$. While the polarization degree from $1.5/\Gamma \leq \theta _{\text{LOS}} \leq 5/\Gamma$ is much smaller. In Figure \ref{fig:higher_PD} (b), we compare the spectra emitted from $0 \leq \theta _{\text{LOS}} \leq 1.5/\Gamma$ and $1.5/\Gamma \leq \theta _{\text{LOS}} \leq 5/\Gamma$. It is found that the spectrum from the inner region ($0 \leq \theta _{\text{LOS}} \leq 1.5/\Gamma$; obtaining higher PD) has higher energy. This may explain the higher PD of AstroSAT (100 keV–600 keV), considering its higher band energy and narrower bandwidth than POLAR (50 keV–500 keV). 

Except for the above origin, there can be two other effects contributing to the higher PD of AstroSAT. First, as stated in \citet{Chat2022}, the narrower bandwidth and higher band energy of AstroSAT may result in shorter burst duration, which prevents the PD decrease originating from spectral overlapping across the burst duration. In \citet{Parsotan2022}, the PD of the time-resolved spectrum is indeed found to be larger than that of the time-integrated spectrum, based on the Monte Carlo simulations for photosphere emission. Second, the distribution for AstroSAT (with higher PD) could also be explained with a larger $\theta_v$ ($\theta_v \gtrsim 0.04 $, see Figure \ref{fig:alpha_PD}(b)), considering the similar slope of $\alpha-$PD distributions for POLAR and AstroSAT. Perhaps, AstroSAT has a lower sensitivity for polarization detections. Then, the high-polarization bursts with larger $\theta_v$ are achieved, leaving the low-polarization bursts unconstrained. In addition to the comparison between POLAR and AstroSAT, we should deal with the high PD values of AstroSAT (all higher than $45\%$). In Equation (\ref{cd}), we adopt the factor of 0.45, corresponding to the PD at a comoving angle of $\pi/2$ and $\tau =1$ in a spherical outflow. However, for the probability photosphere model, the emission from $\tau \gtrsim 1$ also contributes, which could have $\Pi \approx $ 0.6 (see Figure 6 in \citealt{Belo2011}). Also, the structured jet is considered, the $\Pi$ could be much larger due to an increased asymmetry. Notice that the photosphere polarization in the $\gamma$-ray band is claimed to reach $75\%$ or even higher in some intervals, as shown by Figure 2 in \citet{Parsotan2022}.

\section{Summary and discussion}

\label{sec:Conclu}

The radiation mechanism (thermal photosphere or magnetic synchrotron) of GRB prompt emission is tricky to distinguish. Considering the particular jet structure (with angular distribution, especially for the Lorentz factor; \citealt{Pe'er2008,Pe'er2011,Meng2024}) for the photosphere model or the specified magnetic field structure (with radial decay; \citealt{Zhang2011,Geng2018,Yan2024}), the observed GRB spectrum can be reproduced. Meanwhile, the polarization observation is crucial to diagnose the jet structure or the magnetic field configuration. So far, the polarization measurements for the prompt emission of GRBs are reported by the IKAROS-GAP (50 keV–300 keV; \citealt{Yone2011,Yone2012}), the AstroSat-CZTI (100 keV–600 keV; \citealt{Chat2019,Chat2022}), and the POLAR missions (50 keV–500 keV; \citealt{Zhang2019}).

In this work, based on the observational results of POLAR and AstroSAT, we explore the characteristics of the X-ray afterglow shape, the prompt emission ($T_{90}$ vs $L_{\text{iso}}$, $\alpha$) and the polarization degree (polarization degree vs $\alpha$), for the GRBs with reported polarization detection. The following results are revealed. First, the X-ray afterglows show the power-law shape and the $T_{90}\propto (L_{\text{iso}})^{-0.5}$ correlation exists, consistent with the predictions of the classical fireball model and the neutrino annihilation model (NDAF), respectively. Second, the low-energy spectral index $\alpha$ is found to be quite hard, including the GAP sample and especially the GRB 160910A ($\alpha= -0.36$), GRB 170127C ($\alpha= 0.25$) and GRB 180427A ($\alpha= -0.29$). Third, we find the positive correlation of the $\alpha$ and the polarization degree. The bursts with the hardest $\alpha$ (such as GRB 160910A and GRB 170127C) possess the highest polarization degree. Fourth, for the structured jet and large viewing angle, the symmetry break of photon scattering overlapping should invoke certain photosphere polarization. Based on this consideration, adopting $\Gamma \cdot \theta_{c,\Gamma}=1$ and different p values for the jet structure and viewing angle of $\theta _{v}$=0.015,  we well interpret the above positive correlation. With a larger p value, the $\alpha$ is much harder ($\simeq -(1+3/p)/4$), and the polarization degree is much higher due to enhanced asymmetry.

The positive correlation of the $\alpha$ and the polarization degree can be further tested by the future GRB polarization missions, such as LEAP (50 keV–500 keV, NASA Mission; \citealt{McCon2021}) and POLAR-2 \citep{de2022}. POLAR-2 is planned to be deployed on the China Space Station (CSS) in 2026, and consists of three detectors: a Low-energy X-ray Polarization Detector (LPD, 2–10 keV; \citealt{Feng2023,Feng2024,Yi2024}) and a Broad-spectrum Detector (BSD, 8–2000 keV) developed in China, and a High-energy X-ray Polarization Detector (HPD, 30–800 keV; \citealt{Huls2020}) developed in Europe. But two cautions should be taken. First, from our theoretical point, the correlation is valid for relatively small $\theta_{c,\Gamma}$ and moderate viewing angle ($\theta _{v} \lesssim 3 \theta_{c,\Gamma}$), which can be available for most GRBs \citep{Lund2013, Meng2019} and especially for the bright bursts capable of polarization detection. However, with a larger polarization sample, some outliers with large $\theta _{v}$ (much softer $\alpha=-1.5$ and higher polarization degree can be obtained, maybe for GRB
170101A) or large $\theta_{c,\Gamma}$ (with hard $\alpha$ and smaller polarization degree, see Figure \ref{fig:theta_c} (a)) may exist. We can judge the structure condition better by combining the spectral evolutions \citep{ Meng2019}. Second, just like GRB 180103A, some bursts with extremely complex light curve (multiple episodes) can be included. Then, the used $\alpha$ should not correspond to the time-integrated spectrum.

The origin of the GRB polarization is still puzzled, which may be produced by the synchrotron emission with an ordered magnetic field, the photosphere emission, or the Compton drag effect. For GRB 160910A, GRB 170127C and GRB 180427A (with $\alpha \gtrsim -0.36$), and GRB 110721A (with blackbody component), we consider that their polarization is likely to be caused by the photosphere emission with jet asymmetry. In addition to the correlation of the $\alpha$ and the polarization degree for the time-integrated spectrum, we would expand the studies of the photosphere polarization in future works. On the one hand, the time evolution of the polarization properties and comparison with the spectral evolution should be further explored, along with the rotation of the polarization angle (PA) in GRB 170114A and GRB 160821A \citep{Li2024,Wang2024}. On the other hand, the energy dependence of the polarization degree (namely the polarization spectrum) should be investigated in great detail. Then, it can be further checked by the wide-band observation of POLAR-2. 

\begin{acknowledgements}

We thank Bin-Bin Zhang for the useful discussions. We acknowledge the use of the GRB data from \textit{Fermi}/GBM, POLAR, and AstroSAT. Y. Z. M. is supported by the Youth Program of National Natural Science Foundation of China (grant No. 12403045), the Youth Program of Natural Science Foundation of Guangxi (grant No. 2025GXNSFBA069091), and the Starting Foundation of Guangxi University of Science and Technology (grant No. 24Z01). S. Q. Z. is supported by the Youth Program of National Natural Science Foundation of China (grant No. 12503048) and the Starting Foundation of Guangxi University of Science and Technology (grant No. 24Z17). X. F. L. is supported by the National Natural Science Foundation of China (grant No. 12565008).

\end{acknowledgements}

\end{document}